# The scattering phase shifts of the Hulthén-type potential plus Yukawa potential

Oyewumi[*1], K. J. and Oluwadare[+2], O. J

[1]Department of Physics, Federal University of Technology, Minna, Niger State, Nigeria.
[2]Department of Physics, Federal University Oye-Ekiti, Ekiti State, Nigeria.

**Abstract**

The Hulthén- type potential is perturbed by adding Yukawa potential. By applying a short-range approximation to the Schrödinger equation containing this model via standard method, the scattering state solutions and the corresponding phase shifts were obtained. Our numerical calculations and graphical solutions show the dependencies of scattering phase shifts on Yukawa potential constant $A$, asymptotic wave number $k$, screening parameter $\alpha$ and angular momentum quantum number $l$.

**PACS number:** 03.65.-w; 03.65.Nk

**Keywords:** Schrödinger equation; Hulthén- type potential plus Yukawa potential; short-range approximation; scattering states; scattering phase shifts.

## 1. Introduction

The Hulthén type potential plus Yukawa potential defined in this work [1-4] is given by

$$V(r) = -\left[V_o + \frac{A(1-e^{-\alpha r})}{r}\right]\frac{1}{(e^{\alpha r}-1)}, \tag{1}$$

where $\alpha$ is the screening parameter, $V_o$ is the coupling potential strength and $A$ is a constant parameter related to Yukawa potential. The perturbation here is necessary since two models have a wide range of applications in various aspects of physics. The Yukawa potential has previously been used to calculate the energy levels of neutral atoms (see Yahya et al. 2013 [5] and references therein). The Hulthén potential is a useful model that are always attract attentions of researchers in many fields of physics including nuclear and high energy physics (Hulthén & Sugawara 1957), atomic physics [7,8], solid state physics [9]) and chemical physics [10].

The applicability of Hulthén potential led to its modifications by various researchers in relativistic and non-relativistic quantum mechanics [11-19] while different approaches have been employed in studying the above mentioned cases.

[*]On Sabbatical Leave from: Theoretical Physics Section, Department of Physics, University of Ilorin, Ilorin, Nigeria. kjoyewumi66@unilorin.edu.ng  [+]oluwatimilehin.oluwadare@fuoye.edu.ng

To mention a few, Wei et al. [20] obtained the approximate analytical scattering state solutions of the Schrödinger equation with the generalized Hulthén potential for any ℓ-state. Saad [11] studied the bound states of a spinless particle in D-dimensions and found the normalization constant in terms of incomplete Beta function. Also, the scattering state solutions of the Klein-Gordon equation for the Hulthén potential and transmission resonances have been studied by Guo & Fang [21] while the transmission resonances for a Dirac particle in a one dimensional Hulthén potential was presented by Guo et al. [22]. In their work, they presented the reflection and transmission coefficients and the dependency of transmission resonance on the shape of the potential.

However, the scattering state solutions of the Schrödinger equations with Hulthén type potential plus Yukawa potential have not been studied. Therefore, in our own case, we are applying a fundamental idea to obtain the approximate scattering phase shifts and the corresponding bound state energy levels at the real poles S-matrix for the Hulthén-type potential plus Yukawa potential within context of a short range approximation.

Our work is structured as follows: Section 2 contains the basic equations of Schrödinger equation together with the potential and a short range approximation of interest. In section 3, we obtain the scattering state solutions of the Schrödinger equation with Hulthén type potential plus Yukawa potential by applying a short range approximation via standard method. The numerical results and graphical solutions of the wave number dependent-scattering phase shifts are presented in Section 4, while the conclusion is given in Section 5.

## 2. The Basic equations

The Time-independent Schrödinger equation for describing the dynamics of nonrelativistic particles within a physically solvable potential is written as [23]:

$$\left[-\frac{\hbar^2}{2\mu}\nabla^2 + V(r)\right]\Psi(r,\theta,\varphi) = E_{nl}\Psi(r,\theta,\varphi) \quad (2)$$

and defining the wave function $\Psi(r,\theta,\varphi) = r^{-1}R(r)Y(\theta,\varphi)$, the radial part of the Schrödinger equation takes the following form [23]

$$\frac{d^2R_{nl}(r)}{dr^2} + \frac{2\mu}{\hbar^2}\left[E_{nl} - V(r) - \frac{l(l+1)\hbar^2}{2\mu r^2}\right]R_{nl} = 0, \quad (3)$$

where $l$ is the angular momentum quantum number, $\mu$ is the reduced mass of the particles interacting with the potential field $V(r)$ and $E_{nl}$ is the nonrelativistic energy of the particles. By substituting Eq. (1) into Eq. (3), we can simply obtain

$$\frac{d^2 R_{nl}(r)}{dr^2} + \left\{\frac{2\mu E_{nl}}{\hbar^2} + \frac{2\mu}{\hbar^2}\left[V_o + \frac{A(1-e^{-\alpha r})}{r}\right]\frac{1}{(e^{\alpha r}-1)} - \frac{l(l+1)}{r^2}\right\} R_{nl} = 0. \quad (4)$$

To find an approximate solution, we apply short-range approximation of the type [5, 24, 25, 27-29]

$$\frac{1}{r^2} \approx \frac{\alpha^2}{(1-e^{-\alpha r})^2} \quad (5)$$

to remove the effect of the centrifugal term. This approximation is valid only for small values of screening parameter $\alpha$ and fail for large $\alpha$. The effects of a similar short range approximation on the scattering phase shifts have been explained by Oluwadare et al. [24].

### 3. Scattering state solutions

In order to solve for the scattering state solutions, the Schrödinger equation with the Hulthén type potential plus Yukawa potential is then transformed by the variable $z = 1 - e^{-\alpha r}$, which yields

$$(1-z)^2 R_{nl}(z)'' - (1-z) R_{nl}(z)' + z^{-2}[-\zeta_1 z^2 + \zeta_2 z - \zeta_3] R_{nl} = 0, \quad (6)$$

where

$$\zeta_1 = \frac{2\mu}{\alpha^2 \hbar^2}[V_o + \alpha A] - l(l+1) - \frac{k^2}{\alpha^2}, \quad \zeta_2 = \frac{2\mu}{\alpha^2 \hbar^2}[V_o + \alpha A], \quad \zeta_3 = l(l+1), \quad (7)$$

and the propagation constant or the asymptotic wave number (k) of the particles interacting in this potentials equals $\sqrt{\frac{2\mu E_{nl}}{\hbar^2} - \alpha^2 l(l+1)}$. For zero angular momentum quantum number ($l = 0$) in the natural units ($\hbar = \mu = c = 1$) implies that the momentum $k = \sqrt{2E_{nl}}$.

Assuming a useful trial wave function of the form:

$$U(z) = z^\sigma (1-z)^{-ik/\alpha} f(z), \quad (8)$$

with

$$\sigma = \frac{1}{2} + \sqrt{\frac{1}{4} + l(l+1)}. \quad (9)$$

Then, by substituting this into Eq. (6), we have

$$z(1-z)f''(z) + \left[2\sigma - \left(2\sigma - \frac{2ik}{\alpha} + 1\right)z\right]f'(z) + \left[\left(\sigma - \frac{ik}{\alpha}\right)^2 + \zeta_1\right]f(z) = 0. \quad (10)$$

By applying the boundary condition that $f(z)$ tends to a finite as $z \to 0$, the solution [26] is

$$R_{nl}(r) = N_{nl}(1-e^{-\alpha r})^\sigma e^{ikr} {}_2F_1(a; b; c; 1-e^{-\alpha r}), \quad (11)$$

where

$$a = \sigma - \frac{ik}{\alpha} - \sqrt{\zeta_1}, \tag{12a}$$

$$b = \sigma - \frac{ik}{\alpha} + \sqrt{\zeta_1}, \tag{12b}$$

$$c = 2\sigma \tag{12c}$$

and $N_{nl}$ is the normalization constant

### 3.1. Scattering Phase shifts

The scattering phase shifts $\delta_l$ can be obtained by analyzing the asymptotic behaviour of the wave function. Therefore, we consider a recurrence relation of hypergeometric function, which is given by:

$$_2F_1(a; b; c; z) = \frac{\Gamma(c)\Gamma(c-a-b)}{\Gamma(c-a)\Gamma(c-b)} {}_2F_1(a, b; 1+a+b-c; 1-z)$$

$$+ (1-z)^{c-a-b} \frac{\Gamma(c)\Gamma(a+b-c)}{\Gamma(a)\Gamma(b)} {}_2F_1(c-a, c-b; c-a-b+1; 1-z). \tag{13}$$

Using Eq. (13) and the property $_2F_1(a, b; c; 0) = 1$, as $r \to \infty$, we have

$$\lim_{r \to \infty} 2F_1(a, b; c; 1 - e^{-\alpha r}) \sim \left\{ \frac{\Gamma(c-a-b)}{\Gamma(c-a)\,\Gamma(c-b)} + e^{-2ikr} \left[ \frac{\Gamma(c-a-b)}{\Gamma(c-a)\,\Gamma(c-b)} \right]^* \right\}, \tag{14}$$

where we have used the following conjugate relations in the process:

$$(a + b - c)^* = c - a - b = \frac{2ik}{\alpha}, \tag{15a}$$

$$a^* = c - a = \sigma + \frac{ik}{\alpha} + \sqrt{\zeta_1}, \tag{15b}$$

$$b^* = c - b = \sigma + \frac{ik}{\alpha} - \sqrt{\zeta_1}. \tag{15c}$$

A phase parameter $\theta_l$ related to the scattering phase shift is defined by [20, 24]

$$\frac{\Gamma(c-a-b)}{\Gamma(c-a)\,\Gamma(c-b)} = \left| \frac{\Gamma(c-a-b)}{\Gamma(c-a)\,\Gamma(c-b)} \right| e^{i\theta_l}, \tag{16}$$

which is substituted into Eq. (14), leading to

$$\lim_{r \to \infty} 2F_1(a, b; c; 1 - e^{-\alpha r}) \sim e^{-ikr} \left[ e^{i(kr+\theta_l)} + e^{-i(kr+\theta_l)} \right]. \tag{17}$$

Thus, the solution $R(r)$ in Eq. (11) has the asymptotic behaviour of the form;

$$R(r) \sim \cos(kr + \theta_l) r \to \infty, \tag{18}$$

with

$$\theta_l = arg\Gamma\left(\frac{2ik}{\alpha}\right) - arg\Gamma(a^*) - arg\Gamma(b^*) \tag{19}$$

The approximate phase shift $\Delta_l^{(0)}$ without the Hulthén potential plus Yukawa potential is obtained as

$$\theta_l^{(0)} = arg\Gamma\left(\frac{2ik}{\alpha}\right) - arg\Gamma(\eta_1^*) - arg\Gamma(\eta_2^*), \qquad (20)$$

where

$$\eta_1^* = \sigma + \frac{ik}{\alpha} + i\sqrt{l(l+1) + \frac{k^2}{\alpha^2}} \text{ and } \eta_2^* = \sigma + \frac{ik}{\alpha} - i\sqrt{l(l+1) + \frac{k^2}{\alpha^2}}. \qquad (21)$$

A more 'physical' definition of the scattering phase shift $\delta_l$ when applying some short-range type approximations for the angular motions is recently suggested by Oluwadare et al. [24] as

$$\delta_l = \theta_l - \theta_l^{(0)} = arg\Gamma(\eta_1^*) + arg\Gamma(\eta_2^*) - arg\Gamma(a^*) - arg\Gamma(b^*). \qquad (22)$$

### 4.0. Numerical Results

**Table 1:** The wave number-dependent scattering phase shifts $\delta_l$ for a Hulthén type potential plus Yukawa potential in atomic units ($\hbar = \mu = 1$) for $l = 0$ & $V_o = 1$

| k | α | $\delta_l$ for $A = 0$ | $\delta_l$ for $A = 5$ |
|---|---|---|---|
| 0.01 | 0.050 | 85.99747 | 96.02187 |
|  | 0.075 | 56.30153 | 67.31205 |
|  | 0.100 | 42.77292 | 52.53402 |
| 0.03 | 0.050 | 83.01168 | 93.33025 |
|  | 0.075 | 54.93278 | 65.15410 |
|  | 0.100 | 41.11436 | 50.98069 |
| 0.05 | 0.050 | 80.67924 | 90.94696 |
|  | 0.075 | 53.47469 | 63.51416 |
|  | 0.100 | 39.91334 | 49.70670 |
| 0.07 | 0.050 | 78.66066 | 88.84724 |
|  | 0.075 | 52.15129 | 62.09286 |
|  | 0.100 | 38.90094 | 48.61442 |
| 0.09 | 0.050 | 76.85325 | 86.95779 |
|  | 0.075 | 50.95406 | 60.81410 |
|  | 0.100 | 38.00155 | 47.63832 |
| 0.11 | 0.050 | 75.20635 | 85.23028 |
|  | 0.075 | 49.86018 | 59.64271 |
|  | 0.100 | 37.18307 | 46.74557 |
| 0.13 | 0.050 | 73.68827 | 83.63312 |
|  | 0.075 | 48.85074 | 58.55770 |
|  | 0.100 | 36.42819 | 45.91819 |
| 0.15 | 0.050 | 72.27707 | 82.14437 |
|  | 0.075 | 47.91182 | 57.54473 |
|  | 0.100 | 35.72587 | 45.14486 |

**Table 2:** The wave number-dependent scattering phase shifts $\delta_l$ for a Hulthén type potential plus Yukawa potential in atomic units ($\hbar = \mu = 1$) for $l = 1$ & $V_o = 1$

| k | α | $\delta_l$ for $A = 0$ | $\delta_l$ for $A = 5$ |
|---|---|---|---|
| **0.01** | 0.050 | 83.26582 | 93.32007 |
|  | 0.075 | 53.32553 | 64.25479 |
|  | 0.100 | 39.52724 | 49.56731 |
| **0.03** | 0.050 | 80.94760 | 91.28052 |
|  | 0.075 | 52.51662 | 62.75149 |
|  | 0.100 | 38.41452 | 48.39482 |
| **0.05** | 0.050 | 78.95579 | 89.23458 |
|  | 0.075 | 51.43166 | 61.49191 |
|  | 0.100 | 37.59783 | 47.44867 |
| **0.07** | 0.050 | 77.10688 | 87.30387 |
|  | 0.075 | 50.33598 | 60.29911 |
|  | 0.100 | 36.84001 | 46.59532 |
| **0.09** | 0.050 | 75.38692 | 85.50147 |
|  | 0.075 | 49.28127 | 59.16223 |
|  | 0.100 | 36.11447 | 45.78786 |
| **0.11** | 0.050 | 73.78606 | 83.81961 |
|  | 0.075 | 48.27840 | 58.08115 |
|  | 0.100 | 35.41759 | 45.01435 |
| **0.13** | 0.050 | 72.29206 | 82.24618 |
|  | 0.075 | 47.32822 | 57.05467 |
|  | 0.100 | 34.74937 | 44.27207 |
| **0.15** | 0.050 | 70.89283 | 80.76903 |
|  | 0.075 | 46.42846 | 56.08011 |
|  | 0.100 | 34.10972 | 43.56011 |

**Table 3:** The wave number-dependent scattering phase shifts $\delta_l$ for a Hulthén type potential plus Yukawa potential in atomic units ($\hbar = \mu = 1$) for $l = 2$ & $V_o = 1$

| k | α | $\delta_l$ for $A = 0$ | $\delta_l$ for $A = 5$ |
|---|---|---|---|
| **0.01** | 0.050 | 80.11795 | 90.27644 |
|  | 0.075 | 50.09431 | 60.57175 |
|  | 0.100 | 35.20799 | 46.25664 |
| **0.03** | 0.050 | 78.18973 | 88.55603 |
|  | 0.075 | 49.46769 | 59.69517 |
|  | 0.100 | 34.99714 | 45.19103 |
| **0.05** | 0.050 | 76.53055 | 86.83194 |
|  | 0.075 | 48.63354 | 58.72984 |
|  | 0.100 | 34.49899 | 44.45637 |
| **0.07** | 0.050 | 74.93659 | 85.15450 |
|  | 0.075 | 47.76093 | 57.76613 |
|  | 0.100 | 33.95912 | 43.79587 |
| **0.09** | 0.050 | 73.40527 | 83.53989 |
|  | 0.075 | 46.89550 | 56.81843 |
|  | 0.100 | 33.41263 | 43.15913 |
| **0.11** | 0.050 | 71.94191 | 81.99477 |
|  | 0.075 | 46.04966 | 55.89310 |
|  | 0.100 | 32.86989 | 42.53562 |
| **0.13** | 0.050 | 70.54801 | 80.52068 |
|  | 0.075 | 45.22816 | 54.99380 |
|  | 0.100 | 32.33533 | 41.92400 |
| **0.15** | 0.050 | 69.22205 | 79.11607 |
|  | 0.075 | 44.43324 | 54.12254 |
|  | 0.100 | 31.81139 | 41.32505 |

## 4.1. Graphical Results

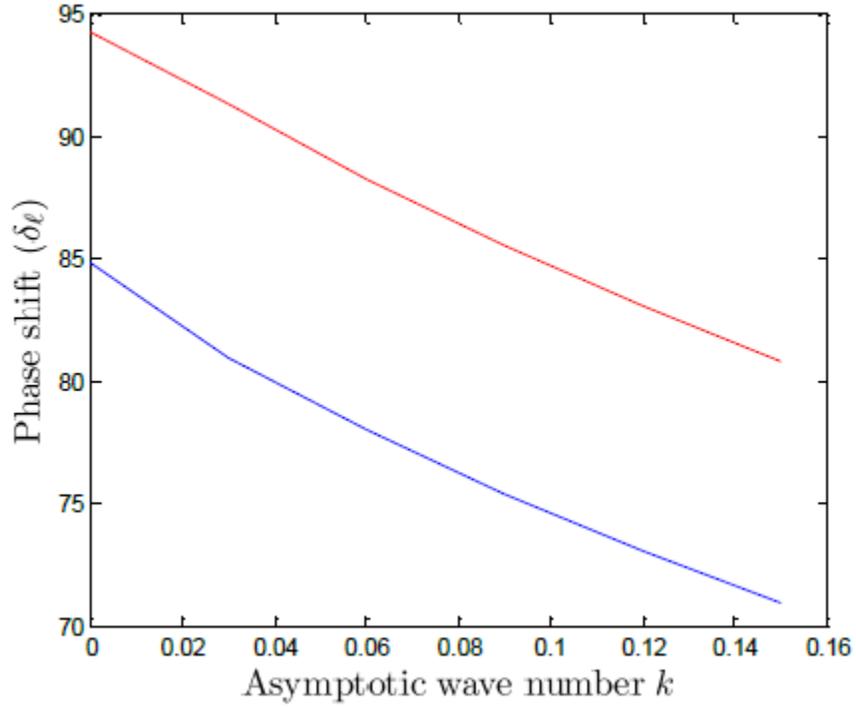

**Fig. 1:** The asymptotic wave number dependent- scattering phase shifts (blue solid line) $l = 1, A = 0, \alpha = 0.050, V_0 = 1$ is compared to that with $l = 1, A = 5, \alpha = 0.050, V_0 = 1$ (red dashed line).

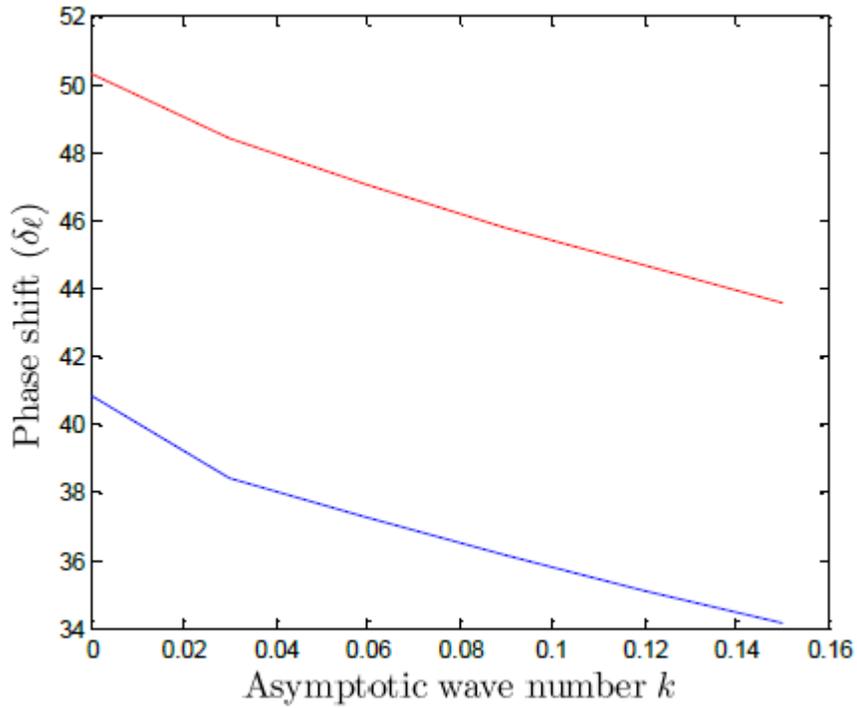

**Fig. 2:** The asymptotic wave number dependent- scattering phase shifts (blue solid line) $l = 1, A = 0, \alpha = 0.100, V_0 = 1$ is compared to that with $l = 1, A = 5, \alpha = 0.100, V_0 = 1$ (red dashed line).

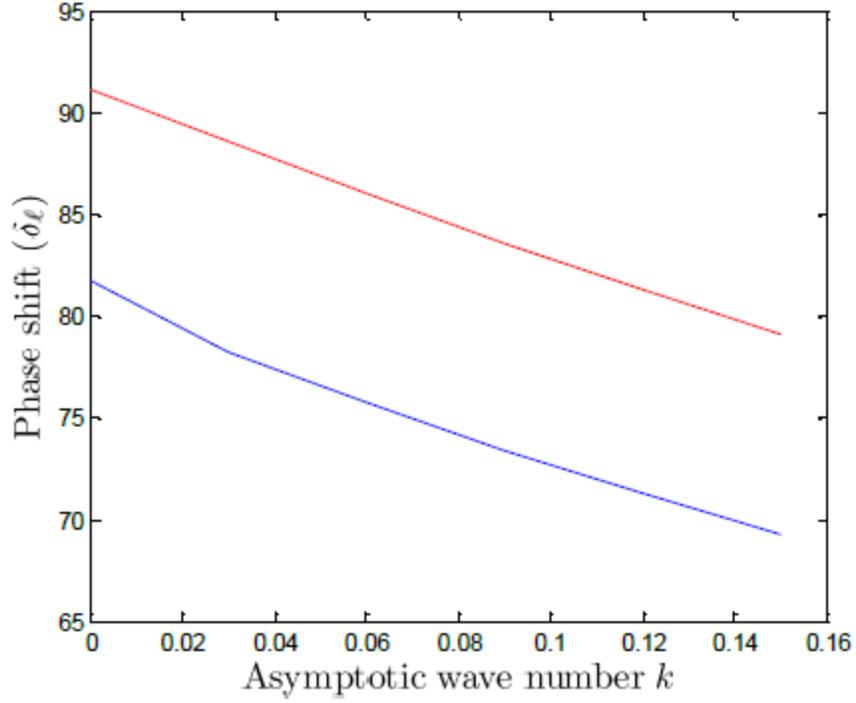

**Fig. 3:** The asymptotic wave number dependent- scattering phase shifts (blue solid line) $l = 2, A = 0, \alpha = 0.050, V_0 = 1$ is compared to that with $l = 2, A = 5, \alpha = 0.050, V_0 = 1$ (red dashed line).

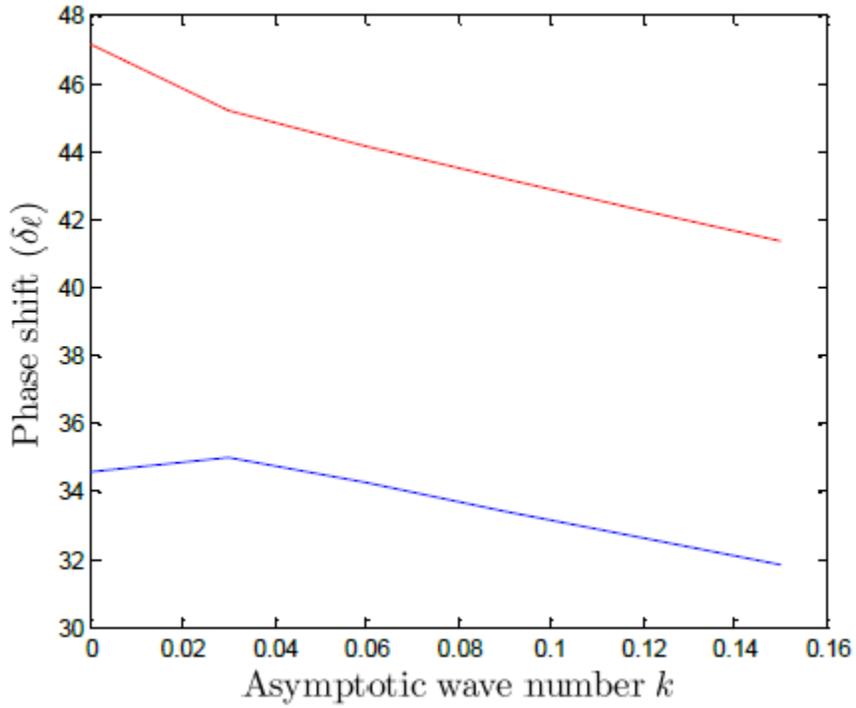

**Fig. 4:** The asymptotic wave number dependent- scattering phase shifts (blue solid line) $l = 2, A = 0, \alpha = 0.100, V_0 = 1$ is compared to that with $l = 2, A = 5, \alpha = 0100, V_0 = 1$ (red dashed line).

**4.2. Discussion**

In order to show the regularity and consistency of our results, we obtained the asymptotic wave number dependent-phase shifts for various values of screening parameter $\alpha = 0.050, 0.075, 0.100$ and Yukawa potential constant $A$.

Table 1, Table 2 & Table 3 show the numerical wave number- dependent phase shifts for $l = 0, 1$ & $2$, respectively. The scattering phase shifts reduce with increasing values of the asymptotic wave number $k$, angular momentum quantum number $l$ and screening parameter $\alpha$ but increase with increasing values of Yukawa potential constant $A$. The results for which the Yukawa potential constant varnishes ($A = 0$) correspond to that of Hulthén-type potential.

The Fig. 1-4 shows the linearity between the phase shifts and asymptotic wave number $k$, angular momentum quantum number $l$ and screening parameter $\alpha$ for the Hulthén potential plus Yukawa potential. Fig. 1-4 also show the effects of Yukawa potential constant $A$ on the asymptotic wave number dependent-scattering phase shifts as the angular momentum quantum number $l$ increases. For $k > 0.03$, the scattering phase shifts reduce linearly.

**5. Conclusion**

When a short range approximation is applied to the Hulthén –type potential plus Yukawa potential with the Schrödinger equation the scattering state solutions of nonrelativistic particles interacting radially with the Hulthén potential plus Yukawa potential were also investigated. The radial scattering wave functions and wave number – dependent phase shifts are presented.

The numerical values of wave number-dependent scattering phase shifts for the Hulthén type potential plus Yukawa potential were displayed in Table 1, Table 2 & Table 3 for $l = 0, 1$ & $2$, respectively. The presence of Yukawa potential increases the values of scattering phase shifts for any arbitrary values of angular momentum quantum number $l$ and the screening parameter $\alpha$. The scattering phase shifts reduces with increasing values of the asymptotic wave number $k$, angular momentum quantum number $l$ and screening parameter while the Fig. 1-4 show the effects of Yukawa potential constant $A$ on the phase shifts and the dependencies of phase shifts on the asymptotic wave number $k$, angular momentum quantum number $l$ and screening parameter $\alpha$.